\documentclass[english,aps,superscriptaddress]{revtex4}
\usepackage[T1]{fontenc}
\usepackage[latin9]{inputenc}
\usepackage{mathrsfs}
\usepackage{amsbsy}
\usepackage{amssymb}
\usepackage{esint}
\usepackage{color}
\makeatletter

\usepackage{babel}
\begin{document}

\title{Constraining non-dissipative transport coefficients in global equilibrium}

\author{Shi-Zheng Yang}

\affiliation{Key Laboratory of Particle Physics and Particle Irradiation (MOE),
Institute of Frontier and Interdisciplinary Science,
Shandong University, Qingdao, Shandong 266237, China}

\author{Jian-Hua Gao}

\affiliation{Shandong Provincial Key Laboratory of Optical Astronomy and Solar-Terrestrial
Environment, Institute of Space Sciences, Shandong University, Weihai,
Shandong 264209, China}

\author{Zuo-Tang Liang}

\affiliation{Key Laboratory of Particle Physics and Particle Irradiation (MOE),
Institute of Frontier and Interdisciplinary Science,
Shandong University, Qingdao, Shandong 266237, China}

\begin{abstract}
The fluid in global equilibrium must fulfill some  constraints. These constraints can be derived from  quantum statistical  theory or kinetic theory. In this paper we will show that how these constraints can be applied to determine the non-dissipative  transport coefficients for chiral systems along with the energy-momentum conservation, chiral anomaly for charge current and  trace anomaly  in energy-momentum tensor.
\end{abstract}
\maketitle

\section{Introduction}
\label{sec:intro}
The  charge currents associated with chiral anomaly exhibit peculiar properties which  normal currents do not possess,
such as the famous  chiral magnetic effect \cite{Vilenkin:1980fu,Fukushima:2008xe}
 and chiral vortical effect\cite{Vilenkin:1978hb,Kharzeev:2007tn,Erdmenger:2008rm}. These currents  are all non-dissipative
and could exist even in global equilibrium. These anomalous currents can be derived from various approaches
such as gauge/gravity duality\cite{Newman:2005hd,Yee:2009vw,Rebhan:2009vc,Gynther:2010ed,Amado:2011zx,Kalaydzhyan:2011vx},
principle of entropy increase \cite{Son:2009tf,Sadofyev:2010pr,Pu:2010as,Kharzeev:2011ds}, Kubo formula from quantum field theory
\cite{Kharzeev:2009pj,Fukushima:2009ft,Landsteiner:2011cp,Hou:2012xg,Lin:2018aon,Feng:2018tpb} and quantum kinetic equation
\cite{Gao:2012ix,Stephanov:2012ki,Manuel:2013zaa,Chen:2014cla,Chen:2015gta,Huang:2018wdl,Gao:2018wmr,Gao:2018jsi}.
In this paper, we would provide another novel method to determine or constrain these non-dissipative  transport coefficients in
 anomalous chiral fluids. Since these currents are non-dissipative,  they could exist even in global equilibrium. However in order to arrive at global equilibrium,  the system  must satisfy  some specific constraints especially when the electromagnetic field is present.  Thanks to these
constraints along with the energy-momentum conservation law,  trace anomaly for energy-momentum tensor and chiral anomaly for charge current, we
can determine or constrain the non-dissipative transport coefficients up to the second order.  In Sec. \ref{sec:ge}, we first review how
the constraint in global equilibrium can be derived from either quantum statistical theory or kinetic theory  when electromagnetic field is imposed.
In Sec.\ref{sec:atc}, we will show how to determine the energy-momentum tensor and charge current from the conservation laws and chiral anomaly.
We summarize our results in Sec.\ref{sec:summary}.

We will choose the metric tensor $g_{\mu\nu}=\mathrm{diag}(1,-1,-1,-1)$
and the Levi-Civita tensor $\varepsilon^{\mu\nu\rho\sigma}$ with the convention $\epsilon^{0123}=1$.
For simplicity, we set the electric charge of the chiral fermion as unit.

\section{Global Equilibrium Constraints}
\label{sec:ge}
When a fluid is in global equilibrium without external fields  \cite{Israel:1976tn,Israel:1979wp}, the fluid four-velocity  $u^\mu$ with $u^2=1$ should  be   expansion-free and shear-free, and that the thermal potential $\bar\mu=\mu/T$ which is defined as the chemical potential $\mu$ divided by the temperature $T$  should be constant, i.e.,
\begin{eqnarray}
\Delta^{\mu\rho}\Delta^{\nu\sigma}\left(\partial_\rho u_\sigma + \partial_\sigma u_\rho\right) = 0 ,\ \ \
\partial_\mu \bar \mu = 0
\end{eqnarray}
where $\Delta^{\mu\nu}= g^{\mu\nu}-u^\mu u^\nu$ denotes the spatial projection tensor. In conjunction with the ideal hydrodynamical equation, it is easy to verify that these above conditions are equivalent to the following equations
\begin{eqnarray}
\partial_\mu \beta_\nu + \partial_\nu \beta_\mu =0  ,\ \ \
\partial_\mu \bar \mu = 0
\end{eqnarray}
where $\beta^\mu = u^\mu/T$ can be referred to as thermal velocity similar to the the thermal potential for chemical potential.
These are just the constraint conditions which should be obeyed by the fluid in global equilibrium without external fields.
When an external electromagnetic field tensor $F_{\mu\nu}$ is present, the constraint conditions are generalized to
\begin{eqnarray}
\label{C-EQ}
\partial_\mu \beta_\nu + \partial_\nu \beta_\mu =0  ,\ \ \
\partial_\mu \bar \mu = -F_{\mu\nu}\beta^\nu
\end{eqnarray}
where the electromagnetic field should be static so as to be able to arrive at the global equilibrium. The second equation above indicates that
the external electromagnetic field is balanced by the gradient of the thermal potential. In this paper, we
will assume further that the electromagnetic field is also homogeneous which means that $F_{\mu\nu}$ must be constant, i.e.,
$\partial_\lambda F_{\mu\nu}=0$.

Now we first  review how these constraint conditions can be derived from more underlying theories.
The derivation from quantum statistical theory is based on global thermodynamic equilibrium density operator which had been
given in details in  \cite{Becattini:2012tc,Buzzegoli:2020ycf}.  The general covariant form of the local thermodynamic equilibrium 
density operator is given by
 \begin{eqnarray}
\hat\rho &=& \frac{1}{Z}\textrm{exp}\left[-\int_\Sigma d\Sigma_\mu \left(\hat T^{\mu\nu}\beta_\nu -\bar \mu \hat j^\mu\right)\right].
\end{eqnarray}
where  $\hat T^{\mu\nu}$ is the symmetric energy-momentum tensor operator, $\hat j^\mu$ the conserved current operator,
 $Z$ is the normalization factor such that $\textrm{tr} \hat \rho =1$, and $\Sigma$ is a spacelike 3-D hypersurface. In global equilibrium, the integrand should be time independent
 \begin{eqnarray}
\int_{\Sigma(\tau)} d\Sigma_\mu \left(\hat T^{\mu\nu}\beta_\nu -\bar \mu \hat j^\mu\right)
-\int_{\Sigma(\tau+\Delta \tau)} d\Sigma_\mu \left(\hat T^{\mu\nu}\beta_\nu -\bar \mu \hat j^\mu\right)=0
\end{eqnarray}
and will not depend on the hypersurface $\Sigma$ any more.
With the  assumption that the field $\beta^\mu$ and $\bar\mu$ vanish at the timelike boundary  which connects two spacelike hypersurface
$\Sigma(\tau)$ and $\Sigma(\tau+\Delta\tau)$ and  according to Gauss's theorem, the above equation implies that the integrand is divergenceless:
\begin{eqnarray}
\partial_\mu \left(\hat T^{\mu\nu}\beta_\nu -\bar \mu \hat j^\mu\right)
=\left(\partial_\mu \hat T^{\mu\nu}\right)\beta_\nu  + \hat T^{\mu\nu}\partial_\mu\beta_\nu
- \left(\partial_\mu \bar \mu\right)  \hat j^\mu -  \bar \mu\left(\partial_\mu  \hat j^\mu\right) =0
\end{eqnarray}
Using  the conservation equations for energy-momentum tensor $\partial_\mu \hat T^{\mu\nu} =  F^{\nu\mu}\hat j_\mu$
and charge current $\partial_\mu  \hat j^\mu =0$
and the fact that the energy-momentum tensor is symmetric, we can obtain
\begin{eqnarray}
\frac{1}{2} \hat T^{\mu\nu} \left(\partial_\mu \beta_\nu + \partial_\nu \beta_\mu\right)
- \left(\partial_\mu \bar \mu + F_{\mu\nu}\beta^\nu \right)  \hat j^\mu  =0
\end{eqnarray}
It is obvious  that this equation always holds if the constraint condition (\ref{C-EQ}) is satisfied.

The global equilibrium condition can also be derived from kinetic theory \cite{Gao:2012ix,Liu:2020flb}.  In equilibrium,
the collision terms  in the  Boltzmann  equation will vanish due to  detailed balancing principle
and the kinetic equation will reduce to  Vlasov equation:
\begin{eqnarray}
\label{Js-eq-0}
\delta(p^2-m^2)  p^\mu \left(\frac{\partial}{\partial x^\mu} - F_{\mu\nu}\frac{\partial}{\partial p_\nu} \right)  f(x,p)  &=& 0\;.
\end{eqnarray}
where $p^\mu$ denotes  four-momentum of the particle with mass  $m$ and we have written Vlasov equation in Lorentz covariant form.
In equilibrium, the distribution function  $f(x,p)$ should depend on $x,p$ through the argument  $\beta\cdot p-\bar\mu$
\begin{eqnarray}
f(x,p)  &=&  g (y),\ \ \ \ y=\beta\cdot p-\bar\mu.
\end{eqnarray}
Then the kinetic equation (\ref{Js-eq-0}) can be expressed as
\begin{eqnarray}
\delta(p^2-m^2)\left[\frac{1}{2}p^\mu p^{\nu}(\partial_{\mu}\beta_{\nu} + \partial_{\nu}\beta_{\mu})
 -p^\mu \partial_{\mu}\bar{\mu}- p^\mu F_{\mu\nu}\beta^\nu  \right]\frac{dg}{dy}
=0 \;, \label{eq:nabla-f}
\end{eqnarray}
Obviously,  the kinetic equation always hold if the equilibrium conditions (\ref{C-EQ}) are satisfied.

 Now let us consider the constraint conditions listed above in more details.  {We can solve the first condition directly
 \cite{Florkowski:2018ahw} and the general solution  is given by
\begin{eqnarray}
\label{beta-mu}
\beta_\mu &=&b_\mu -\Omega_{\mu\nu}x^\nu
\end{eqnarray}
where $b_\mu $ is a constant vector and $\Omega_{\mu\nu}$ is a constant antisymmetric tensor. Actually
$\Omega_{\mu\nu}$ is just the thermal vorticity tensor of the fluid (there is a minus sign difference from usual definition)
\begin{eqnarray}
\Omega_{\mu\nu}=\frac{1}{2}\left(\partial_\mu \beta_\nu -\partial_\nu \beta_\mu\right)\;.
\end{eqnarray}
The second condition in (\ref{C-EQ}) has a solution only if the integrability condition is fulfilled\cite{Yang:2020mtz}. It can be obtained
 by differentiating both sides of second equation in  Eq.(\ref{C-EQ}) with $\partial_\nu$  and using the
commutating property of ordinary partial derivatives
\begin{eqnarray}
\label{pd3-a}
\partial_\nu \partial_\mu \bar\mu =\partial_\mu \partial_\nu \bar \mu =  - F_{\mu\lambda}\partial_\nu  \beta^\lambda
= -F_{\nu\lambda}\partial_\mu  \beta^\lambda ,
\end{eqnarray}
Together  with  Eq.(\ref{beta-mu}), the above equation can written as
\begin{eqnarray}
\label{integrability}
{F_{\lambda}}^\mu \Omega^{\nu\lambda}-{F_{\lambda}}^\nu \Omega^{\mu\lambda}=0\;,
\end{eqnarray}
The general solution under this integrability condition is given by
\begin{eqnarray}
\bar\mu &=&-\frac{1}{2}F^{\mu\lambda} x_\lambda \Omega_{\mu \nu} x^\nu + c
\end{eqnarray}
We can decompose the antisymmetric tensors $F_{\mu\nu} $ and $\Omega_{\mu\nu}$ with the fluid velocity $u_\mu $ as
\begin{eqnarray}
\label{FEB1}
F_{\mu\nu}&=& E_\mu u_\nu -E_\nu u_\mu + \epsilon_{\mu\nu\rho\sigma}u^\rho B^\sigma \;,\\
\label{VEB1}
\Omega_{\mu\nu}&=&\frac{1}{T} \left(\varepsilon_\mu u_\nu -\varepsilon_\nu u_\mu
+ \epsilon_{\mu\nu\rho\sigma}u^\rho \omega^\sigma\right) \;,
\end{eqnarray}
where the  electric field $E^\mu$, magnetic field $B^\mu$, acceleration vector $\varepsilon^\mu$ and vorticity vector $\omega^\mu$
are given by,respectively,
\begin{eqnarray}
E^\mu &=& F^{\mu\nu}u_\nu \;,\ \ \ \ \ B^\mu = \frac{1}{2}\epsilon^{\mu\nu\alpha\beta}u_\nu F_{\alpha\beta}\;,\\
\varepsilon^\mu &=& T\Omega^{\mu\nu}u_\nu \;,\ \ \ \omega^\mu =\frac{1}{2}\epsilon^{\mu\nu\alpha\beta}u_\nu \partial_\alpha^x u_\beta \;.
\end{eqnarray}
With this decomposition, it is easy to verify that  the integrability  condition (\ref{integrability})  is equivalent to
\begin{eqnarray}
\label{integrability-1}
E^\mu\omega^\nu -E^\nu\omega^\mu = -  B^\mu \varepsilon^\nu +  B^\nu \varepsilon^\mu ,\ \ \ \
E^\mu \varepsilon^\nu - E^\nu \varepsilon^\mu =B^\mu \omega^\nu  -  B^\nu\omega^\mu   \;.
\end{eqnarray}
We will show that these  relations  play an important role to determine  the possible forms of the non-dissipative terms in
energy-momentum tensor and charge current in global equilibrium.
\section{Non-dissipative Transport Coefficients}
\label{sec:atc}
In this section, we will apply the conservation laws and trace anomaly to constrain the possible anomalous transport coefficients in a chiral system
in which only right-hand or left-hand Weyl fermions are involved.  These conservation laws and trace anomaly are given by
\begin{eqnarray}
\label{conservation}
\partial_\mu  T ^{\mu\nu} =  F^{\nu\mu}{ j}_\mu ,\ \ \ \  \partial_\mu   j^\mu  = C  E\cdot B,\ \ \ \ g_{\mu\nu} T^{\mu\nu}= \tilde C (E^2-B^2)
\end{eqnarray}
We will expand the energy-momentum tensor and charge current in powers of $F^{\mu\nu}$ and $\Omega^{\mu\nu}$ or equivalently in terms of $B^\mu$,$E^\mu$, $\omega^\mu$ and $\varepsilon^\mu$. Since $F^{\mu\nu}$ and $\Omega^{\mu\nu}$ are both constant, it is unnecessary to  consider $\partial_\mu T$, $\partial_\mu u_\nu$ and  $\partial_\mu \bar \mu$  because all these derivatives  can be  expressed as the linear combination of $E^\mu$, $\omega^\mu$ and $\varepsilon^\mu$  by using the constraint condition (\ref{C-EQ}), e.g.,
 \begin{eqnarray}
 \label{relations}
\partial_\mu T =- T \varepsilon_\mu, \ \ \ \ \
\partial_\mu u_\nu = -u_\mu \varepsilon_\nu  +\epsilon_{\mu\nu\alpha\beta}u^\alpha \omega^\beta,\ \ \ \
\partial_\mu \bar \mu  = -\frac{E_\mu}{T}
\end{eqnarray}
We take $u^\mu$, $T$ and $\bar \mu$ to be of the zeroth order, $F^{\mu\nu}$ and $\Omega^{\mu\nu}$ to be of the  first order and so on.

Let us start with the zeroth-order $T ^{\mu\nu}$ and $j ^\mu$. They are just the well-known ideal hydrodynamical results:
\begin{eqnarray}
 T ^{(0)\mu\nu} = \rho  u^\mu u^\nu -P  \Delta^{\mu\nu}, \ \ \ j ^{(0)\mu} = n  u^\mu
\end{eqnarray}
where $\rho $ is the energy density, $P $ the pressure and $n $  the charge density. It is easy to verify that
\begin{eqnarray}
\partial_\mu  T ^{(0)\mu\nu} = (\rho + P)  u^\mu \partial_\mu u^\nu - \partial^\nu P
= -\rho\varepsilon^\nu - T \partial \frac{P}{T}
\end{eqnarray}
Using  the thermal identity
\begin{eqnarray}
d\frac{P }{T}=n  d\bar\mu  -\rho  d \frac{1}{T}
\end{eqnarray}
and the last equation in (\ref{relations}), we obtain
\begin{eqnarray}
\partial_\mu  T ^{(0)\mu\nu} = E^\nu n  = F^{\nu\mu}j_{s\mu}^{(0)}
\end{eqnarray}
which indicates that  the energy-momentum conservation law holds automatically.
It is trivial to show that at zeroth-order charge current is also conserved automatically
\begin{eqnarray}
\partial_\mu   j^{(0)\mu}  = 0
\end{eqnarray}
There is no chiral anomaly at zeroth order as it should be. For the massless fermions, the conformal symmetry holds at the zeroth order and
the trace of energy-momentum tensor must vanish which results in the well-known relation
\begin{eqnarray}
 \rho= 3P.
\end{eqnarray}

  When we go beyond the zeroth-order, we need first pin down which frame we choose for the fluid velocity $u^\mu$. In our work,
we will use the  $\beta$ frame introduced in \cite{Becattini:2014yxa}, In this frame, the non-dissipative  coefficients in
global equilibrium would take more elegant form.  We will assume the interactions which controls the chiral system keep
charge, parity and time reversal invariance. Then at first order, the general expressions for the energy-momentum tensor and charge current
take the following form
\begin{eqnarray}
 T ^{(1)\mu\nu} &=& \lambda ^\omega\left(u^\mu \omega^\nu + u^\nu \omega^\mu  \right)
 + \lambda ^B \left(u^\mu B^\nu + u^\nu B^\mu  \right),\\
 j ^{(1)\mu} &=& \xi  \omega^\mu + \xi ^B B^\mu
\end{eqnarray}
With this expression, the divergence of the  current reads
\begin{eqnarray}
\label{j-s-1}
\partial_\mu j ^{(1)\mu} &=&
\frac{\partial \xi }{\partial T} \partial_\mu T  \omega^\mu
+\frac{\partial \xi }{\partial \bar \mu } \partial_\mu \bar\mu   \omega^\mu
+ \xi  \partial_\mu \omega^\mu
+\frac{\partial \xi^B }{\partial T} \partial_\mu T  B^\mu
+\frac{\partial \xi^B }{\partial \bar \mu } \partial_\mu \bar\mu   B^\mu
+ \xi^B  \partial_\mu B^\mu
\end{eqnarray}
Using the relations (\ref{relations}) and the derived relations below
\begin{eqnarray}
\partial_\mu \omega_\nu &=&  \varepsilon\cdot \omega\, g_{\mu\nu} - 2\varepsilon_\mu \omega_\nu,\\
\partial_\mu B_\nu &=&-  E_\mu \omega_\nu
 + \varepsilon\cdot B\, u_\mu u_\nu + \omega\cdot E\, \Delta_{\mu\nu}
-\left(u_\mu \epsilon_{\nu\lambda\rho\sigma}+ u_\nu \epsilon_{\mu\lambda\rho\sigma}\right)
u^\lambda \varepsilon^\rho E^\sigma
\end{eqnarray}
the equation (\ref{j-s-1}) can be written as
\begin{eqnarray}
\label{j-s-1-a}
\partial_\mu j ^{(1)\mu} &=& \left(2\xi  - T\frac{\partial \xi }{\partial T}\right)\varepsilon \cdot \omega
+\left(2 \xi ^B - \frac{1}{T}\frac{\partial \xi }{\partial \bar \mu } \right) E\cdot \omega
+ \left(\xi ^B - T\frac{\partial \xi^B }{\partial T}\right) \varepsilon\cdot B
-\frac{1}{T}\frac{\partial \xi^B }{\partial \bar \mu } E\cdot B
\end{eqnarray}
The fact that this result should equal to the anomalous term $C  E\cdot B$ from the second equation in (\ref{conservation})
lead to the following equations:
\begin{eqnarray}
2\xi  - T\frac{\partial \xi }{\partial T} = 0 ,\ \ \
 2 \xi ^B - \frac{1}{T}\frac{\partial \xi }{\partial \bar \mu } = 0,\ \ \
\xi ^B - T\frac{\partial \xi^B }{\partial T}= 0 ,\ \ \
-\frac{1}{T}\frac{\partial \xi^B }{\partial \bar \mu }= C .
\end{eqnarray}
The general solution for this set of equations are easy to obtain
\begin{eqnarray}
\label{xi-B-s}
\xi^B  &=& -C  T \bar\mu   + b T= -C  \mu  + b T,\\
\label{xi-s}
\xi  &=& -C  T^2 \bar\mu ^2  + 2 b T^2\bar \mu  + a T^2 =-C  \mu ^2 + 2 b T \mu  + a T^2
\end{eqnarray}
where $a$ and $b$ are both integral constants. It should be noted that the temperature dependence derived from the differential equations
are consistent with the direct dimension analysis. Actually it is more convenient to determine the temperature power from dimension analysis.
These results had  been derived from the anomalous hydrodynamics by using the principle of entropy increase\cite{Son:2009tf,Sadofyev:2010pr,Pu:2010as,Kharzeev:2011ds}. However it seems as if our method given here
involve much less calculations.  Similarly, the divergence of the energy-momentum tensor can be expressed as
\begin{eqnarray}
\partial_\mu  T ^{(1)\mu\nu} &=&
\frac{\partial \lambda }{\partial T} \partial_\mu T \left( u^\mu \omega^\nu + u^\nu \omega^\mu\right)
+\frac{\partial \lambda }{\partial \bar \mu } \partial_\mu \bar\mu   \left( u^\mu \omega^\nu + u^\nu \omega^\mu\right)
+ \lambda  \partial_\mu \left( u^\mu \omega^\nu + u^\nu \omega^\mu\right)\nonumber\\
& &+\frac{\partial \lambda^B }{\partial T} \partial_\mu T  \left( u^\mu B^\nu + u^\nu B^\mu\right)
+\frac{\partial \lambda^B }{\partial \bar \mu } \partial_\mu \bar\mu    \left( u^\mu B^\nu + u^\nu B^\mu\right)
+ \lambda^B  \partial_\mu  \left( u^\mu B^\nu + u^\nu B^\mu\right)\nonumber\\
&=&\left[ (3\lambda  - T\frac{\partial \lambda }{\partial T})\varepsilon \cdot \omega
+(2 \lambda ^B - \frac{1}{T}\frac{\partial \lambda }{\partial \bar \mu }) E\cdot \omega
+ (2\lambda ^B - T\frac{\partial \lambda^B }{\partial T}) \varepsilon\cdot B
-\frac{1}{T}\frac{\partial \lambda^B }{\partial \bar \mu } E\cdot B\right] u^\nu\nonumber\\
& & - 2 \lambda ^B \epsilon^{\nu\alpha\beta\gamma}u_\alpha \omega_\beta B_\gamma
\end{eqnarray}
where we have used the second identity in Eq.(\ref{integrability-1}).
The righthand of the energy-momentum conservation at first order in Eq.(\ref{conservation}) is given by
\begin{eqnarray}
F^{\nu\mu}j_{s\mu}^{(1)}&=& -\xi  (E\cdot \omega) u_\nu - \xi ^B (E\cdot B) u^\nu
- \xi  \epsilon^{\nu\alpha\beta\gamma}u_\alpha \omega_\beta B_\gamma
\end{eqnarray}
Then the conservation law $\partial_\mu  T ^{(1)\mu\nu} = F^{\nu\mu}j_{s\mu}^{(1)}$ requires
\begin{eqnarray}
3\lambda  - T\frac{\partial \lambda }{\partial T}=0,\ \ \
2 \lambda ^B - \frac{1}{T}\frac{\partial \lambda }{\partial \bar \mu } = -\xi ,\ \ \
2\lambda ^B - T\frac{\partial \lambda^B }{\partial T}= 0,\ \ \
\frac{1}{T}\frac{\partial \lambda^B }{\partial \bar \mu } = \xi ^B,\ \ \
2 \lambda ^B =  \xi
\end{eqnarray}
From the last equation, we note that the coefficient $\lambda ^B$ has been totally determined by the coefficient $\xi $ in the charge current.
It is trivial to verify  that both  second and  third   last equations hold automatically with the result of $\xi $ in Eq.(\ref{xi-s}).
Substituting the result of $\lambda ^B$ into the first and second equations, we can obtain the general expression for $\lambda $. We list the
solution for $\lambda ^B $ and $\lambda $ in the following:
\begin{eqnarray}
\lambda ^B &=& \frac{1}{2} \xi  = \frac{1}{2}\left(-C   \bar\mu ^2  + 2 b \bar \mu  + a \right)T^2,\\
\lambda  &=& \frac{2}{3}\left(-C   \bar\mu ^3  + 3 b \bar \mu ^2 + a \bar\mu  + c\right)T^3
\end{eqnarray}
where $c$ is another integral constant. Similarly, the temperature dependence can also be obtained from direct dimension analysis.
It is obvious that  energy-momentum tensor at first order is traceless automatically.

Now let us move on to consider the second-order case. The charge current and energy-momentum tensor at second order take the general form
\begin{eqnarray}
j^{(2)\mu}  &=&\left(\xi ^{\varepsilon\varepsilon} \varepsilon^2+\xi ^{\omega\omega}\omega^2
+\xi ^{\varepsilon E}  \varepsilon\cdot E + \xi ^{\omega B} \omega\cdot B
+\xi ^{E E}  E^2 + \xi ^{B B} B^2 \right)u^\mu  \nonumber\\
& &+ \xi ^{\varepsilon \omega} \epsilon^{\mu \nu \rho\sigma} u_\nu \varepsilon_\rho \omega_\sigma
+ \xi ^{\omega E} \epsilon^{\mu \nu \rho\sigma} u_\nu E_\rho \omega_\sigma
+  \xi ^{E B} \epsilon^{\mu\nu\rho\sigma} u_\nu  E_\rho B_\sigma,\\
T_{s}^{(2)\mu\nu}&=&\left(\lambda ^{\varepsilon\varepsilon} \varepsilon^2+\lambda ^{\omega\omega}\omega^2
+\lambda ^{\varepsilon E}  \varepsilon\cdot E + \lambda ^{\omega B} \omega\cdot B
+\lambda ^{E E}  E^2 + \lambda ^{B B} B^2 \right)u^\mu u^\nu\nonumber\\
& &+ \left(\bar\lambda ^{\varepsilon\varepsilon} \varepsilon^2+\bar\lambda ^{\omega\omega}\omega^2
+\bar\lambda ^{\varepsilon E}  \varepsilon\cdot E + \bar\lambda ^{\omega B} \omega\cdot B
+\bar\lambda ^{E E}  E^2 + \bar\lambda ^{B B} B^2 \right)\Delta^{\mu\nu}\nonumber\\
& &+ \tilde\lambda ^{\varepsilon\varepsilon} \varepsilon^\mu \varepsilon^\nu
+\tilde\lambda ^{\omega\omega}\omega^\mu \omega^\nu
+\tilde\lambda ^{\varepsilon E}\left( \varepsilon^\mu E^\nu + \varepsilon^\nu E^\mu \right)\nonumber\\
& &+ \tilde\lambda ^{\omega B}\left( \omega^\mu  B^\nu + \omega^\nu  B^\mu\right)
+\tilde\lambda ^{E E} E^\mu E^\nu  + \bar\lambda ^{B B} B^\mu B^\nu \nonumber\\
& &+ ( u^\mu \epsilon^{\nu\alpha\beta\gamma} + u^\nu \epsilon^{\mu\alpha\beta\gamma} )
u_\alpha \left(\lambda ^{\varepsilon \omega}\varepsilon_\beta \omega_\gamma
+ \lambda ^{\omega E} E_\beta \omega_\gamma + \lambda ^{E B}E_\beta B_\gamma\right)
\end{eqnarray}
In order to calculate the divergence of these quantities, we need other useful relations:
\begin{eqnarray}
\partial_\mu \varepsilon_\nu &=& \omega_\mu \omega_\nu -\varepsilon_\mu \varepsilon_\nu + \varepsilon^2 u_\mu u_\nu
-\omega^2  \Delta_{\mu\nu}+ \left(u_\mu\epsilon_{\nu\lambda\rho\sigma}+u_\nu \epsilon_{\mu\lambda\rho\sigma}\right)
u^\lambda \varepsilon^\rho \omega^\sigma,\\
\partial_\mu E_\nu &=&  B_\mu \omega_\nu
+ \varepsilon\cdot E\, u_\mu u_\nu - \omega\cdot B\, \Delta_{\mu\nu}
+\left(u_\mu \epsilon_{\nu\lambda\rho\sigma}+ u_\nu \epsilon_{\mu\lambda\rho\sigma}\right)
u^\lambda  E^\rho \omega^\sigma,\\
 0 &=&\epsilon^{\mu\alpha\beta\gamma} \varepsilon_\alpha \omega_\beta E_\gamma
=\epsilon^{\mu\alpha\beta\gamma} \omega_\alpha E_\beta B_\gamma
=\epsilon^{\mu\alpha\beta\gamma}E_\alpha B_\beta \varepsilon_\gamma
=\epsilon^{\mu\alpha\beta\gamma}B_\alpha \varepsilon_\beta \omega_\gamma
\end{eqnarray}
All these relations  can be derived from the first-order relations \ref{relations}. It is easy to verify that the conservation law for the  charge current $\partial_\mu j^{(2)\mu}  =0$ is satisfied automatically. Although we can not
constrain any  coefficients appearing in the sencond-order current $j^{(2)\mu} $, we still can relate the coefficients in second-order energy-momentum tensor $T_{s}^{(2)\mu\nu}$ to the ones in $j^{(2)\mu} $ through the energy-momentum conservation.
Following the same step as we did at first order, the divergence of the energy-momentum tensor reads
\begin{eqnarray}
\label{dT-2}
\partial_\mu  T ^{(2)\mu\nu}
&=&\mathcal{X}_1 \varepsilon^2 \varepsilon^\nu
+\mathcal{X}_2\omega^2 \varepsilon^\nu
+\mathcal{X}_3 \varepsilon\cdot \omega \omega^\nu \nonumber\\
& &+\mathcal{X}_4 \omega\cdot B \varepsilon^\nu
+\mathcal{X}_5 \varepsilon\cdot B \omega^\nu
+\mathcal{X}_6\omega\cdot E \omega^\nu
+\mathcal{X}_7 \varepsilon^2 E^\nu
+\mathcal{X}_8 \omega^2 E^\nu \nonumber\\
& &+\mathcal{X}_{9} E^2  \varepsilon^\nu
+\mathcal{X}_{10}B^2 \varepsilon^\nu
+\mathcal{X}_{11}E\cdot B\omega^\nu
+\mathcal{X}_{12} \varepsilon\cdot E E^\nu +\mathcal{X}_{13}\omega\cdot B E^\nu \nonumber\\
& &+\mathcal{X}_{14} E^2 E^\nu
+\mathcal{X}_{15} B^2 E^\nu
+\mathcal{X}_{16}E\cdot B B^\nu
\end{eqnarray}
where the coefficients $\mathcal{X}_1$, $\mathcal{X}_2$ and $\mathcal{X}_3$ which are irrelevant to electromagnetic field reads
\begin{eqnarray}
\mathcal{X}_1
&=&
-\lambda ^{\varepsilon\varepsilon}-\bar\lambda ^{\varepsilon\varepsilon}
-\tilde\lambda ^{\varepsilon\varepsilon}-T\frac{\partial \bar\lambda ^{\varepsilon\varepsilon}}{\partial T}
-T\frac{\partial \tilde\lambda ^{\varepsilon\varepsilon}}{\partial T},
\nonumber\\
\mathcal{X}_2
&=&
-\lambda ^{\omega\omega} - 2 \bar\lambda ^{\varepsilon\varepsilon}
+ {2}\lambda ^{\varepsilon \omega}
-3\bar\lambda ^{\omega\omega} - 3  \tilde\lambda ^{\varepsilon\varepsilon}
- T\frac{\partial \bar\lambda ^{\omega\omega}}{\partial T},
\nonumber\\
\mathcal{X}_3
&=&2 \bar\lambda ^{\varepsilon\varepsilon} + 2\bar\lambda ^{\omega\omega}
+ \tilde\lambda ^{\varepsilon\varepsilon} +\tilde\lambda ^{\omega\omega}
- {2}\lambda ^{\varepsilon \omega}-T\frac{\partial\tilde\lambda ^{\omega\omega}}{\partial T},
\end{eqnarray}
the coefficients from $\mathcal{X}_4$ to $\mathcal{X}_8$ with   linear dependence on electromagnetic field are  given by
\begin{eqnarray}
\mathcal{X}_4
&=& \lambda ^{\varepsilon E}-3\tilde\lambda ^{\varepsilon E}
 - \lambda ^{\omega B} +3\tilde\lambda ^{\omega B}
+T\frac{\partial \bar\lambda ^{\varepsilon E}}{\partial T}
+ T\frac{\partial\tilde\lambda ^{\varepsilon E}}{\partial T}
 -T\frac{\partial \bar\lambda ^{\omega B}}{\partial T}
-T\frac{\partial \tilde\lambda ^{\omega B}}{\partial T}
+\frac{1}{T} \frac{\partial \tilde\lambda ^{\varepsilon\varepsilon} }{\partial \bar\mu }
\nonumber\\
\mathcal{X}_5
&=&- \lambda ^{\varepsilon E} +\tilde\lambda ^{\varepsilon E}
-\tilde\lambda ^{\omega B}
-T\frac{\partial \bar\lambda ^{\varepsilon E}}{\partial T}
- T\frac{\partial\tilde\lambda ^{\varepsilon E}}{\partial T}
-T\frac{\partial \tilde\lambda ^{\omega B}}{\partial T}
-\frac{1}{T} \frac{\partial \tilde\lambda ^{\varepsilon\varepsilon} }{\partial \bar\mu }
\nonumber\\
\mathcal{X}_6
&=& \lambda ^{\varepsilon E}+2\bar\lambda ^{\varepsilon E} +\tilde\lambda ^{\varepsilon E}
 + 2 \bar\lambda ^{\omega B}+5\tilde\lambda ^{\omega B} - {2}\lambda ^{\omega E} \nonumber\\
& &+T\frac{\partial \bar\lambda ^{\varepsilon E}}{\partial T}
+ T\frac{\partial\tilde\lambda ^{\varepsilon E}}{\partial T}
-T\frac{\partial \tilde\lambda ^{\omega B}}{\partial T}
+\frac{1}{T} \frac{\partial \tilde\lambda ^{\varepsilon\varepsilon} }{\partial \bar\mu }
-\frac{1}{T}\frac{\partial\tilde\lambda ^{\omega\omega}}{\partial \bar\mu }
\nonumber\\
\mathcal{X}_7
&=&- \lambda ^{\varepsilon E}
-T\frac{\partial \bar\lambda ^{\varepsilon E}}{\partial T}
-2 T\frac{\partial\tilde\lambda ^{\varepsilon E}}{\partial T}
-\frac{1}{T}\frac{\partial \bar\lambda ^{\varepsilon\varepsilon}}{\partial \bar \mu }
-\frac{1}{T} \frac{\partial \tilde\lambda ^{\varepsilon\varepsilon} }{\partial \bar\mu }
\nonumber\\
\mathcal{X}_8
&=&- \lambda ^{\varepsilon E}-2\bar\lambda ^{\varepsilon E}
-3\tilde\lambda ^{\varepsilon E}-2 \bar\lambda ^{\omega B}-3\tilde\lambda ^{\omega B}
+ {2} \lambda ^{\omega E}\nonumber\\
& &-T\frac{\partial \bar\lambda ^{\varepsilon E}}{\partial T}
- T\frac{\partial\tilde\lambda ^{\varepsilon E}}{\partial T}
+T\frac{\partial \tilde\lambda ^{\omega B}}{\partial T}
-\frac{1}{T} \frac{\partial \tilde\lambda ^{\varepsilon\varepsilon} }{\partial \bar\mu }
-\frac{1}{T}\frac{\partial \bar\lambda ^{\omega\omega}}{\partial \bar\mu }
\end{eqnarray}
the coefficients with  double linear dependence on  electromagnetic field are
\begin{eqnarray}
\mathcal{X}_{9}
&=&
-\bar\lambda ^{E E} - \lambda ^{E E}-2 \bar\lambda ^{B B} -3\tilde\lambda ^{B B}
+ {2} \lambda ^{E B}-T \frac{\partial \bar\lambda ^{E E}}{\partial T}
- \frac{1}{T}\frac{\partial\tilde\lambda ^{\varepsilon E}}{\partial \bar\mu }
+\frac{1}{T} \frac{\partial \tilde\lambda ^{\omega B}}{\partial \bar\mu }
\nonumber\\
\mathcal{X}_{10}
&=&
\bar\lambda ^{B B} + \tilde\lambda ^{B B} -  \lambda ^{B B}
-T\frac{\partial \bar\lambda ^{B B}}{\partial T}-T\frac{\partial\tilde\lambda ^{BB}}{\partial T}
\nonumber\\
\mathcal{X}_{11}
&=&
2\bar\lambda ^{E E}+2 \bar\lambda ^{B B}
+\tilde\lambda ^{E E} + 3 \tilde\lambda ^{B B} - {2} \lambda ^{E B}
-T\frac{\partial\tilde\lambda ^{BB}}{\partial T}
-\frac{2}{T} \frac{\partial \tilde\lambda ^{\omega B}}{\partial \bar\mu }
\nonumber\\
\mathcal{X}_{12}
&=&
2\bar\lambda ^{E E} +2 \bar\lambda ^{B B}
+\tilde\lambda ^{E E}+3\tilde\lambda ^{B B}-  {2}\lambda ^{E B}
- \frac{1}{T}\frac{\partial \bar\lambda ^{\varepsilon E}}{\partial \bar \mu }
- \frac{1}{T}\frac{\partial\tilde\lambda ^{\varepsilon E}}{\partial \bar\mu }
-\frac{1}{T} \frac{\partial \tilde\lambda ^{\omega B}}{\partial \bar\mu }
-T\frac{\partial\tilde\lambda ^{E E}}{\partial T}
\nonumber\\
\mathcal{X}_{13}
&=&
 {2} \lambda ^{E B}  {- 2\bar\lambda ^{E E} - 2\bar\lambda ^{B B}}
 -3\tilde\lambda ^{E E} - \tilde\lambda ^{B B}
+T\frac{\partial\tilde\lambda ^{BB}}{\partial T}
- \frac{1}{T} \frac{\partial \bar\lambda ^{\omega B}}{\partial \bar\mu }
\end{eqnarray}
and the coefficients with triple linear dependence on electromagnetic field are given by
\begin{eqnarray}
\mathcal{X}_{14}
=-\frac{1}{T}\left(\frac{\partial \bar\lambda ^{E E}}{\partial \bar \mu }
+\frac{\partial\tilde\lambda ^{E E}}{\partial \bar\mu }\right),\ \ \
\mathcal{X}_{15}
= {-}\frac{1}{T}\frac{\partial \bar\lambda ^{B B}}{\partial \bar \mu } ,\ \ \
\mathcal{X}_{16}=
 -\frac{1}{T}\frac{\partial\tilde\lambda ^{BB}}{\partial \bar\mu }E\cdot B B^\nu
\end{eqnarray}
It should be noted that  in order to arrive at the final result above (\ref{dT-2}), we have used the following identities
\begin{eqnarray}
\varepsilon\cdot \omega B^\nu &=& \varepsilon\cdot B \omega^\nu  + \varepsilon^2 E^\nu - \varepsilon\cdot E \varepsilon^\nu,\nonumber\\
\omega\cdot E B^\nu &=& E\cdot B \omega^\nu + \varepsilon\cdot E E^\nu -E^2 \varepsilon^\nu,\nonumber\\
\varepsilon\cdot B B^\nu &=& B^2\varepsilon^\nu - \omega\cdot B E^\nu + E\cdot B \omega^\nu,\nonumber\\
\varepsilon\cdot E \varepsilon^\nu &=& \varepsilon \cdot B \omega^\nu + \varepsilon^2 E^\nu + \omega^2 E^\nu
- \omega\cdot B \varepsilon^\nu -\omega\cdot E \omega^\nu
\end{eqnarray}
which can be derived directly from the constraint (\ref{integrability-1}). With these identities, we  express the final result as the linear combination of independent terms. The source contribution from the coupling between the electromagnetic field and charge current is
given by
 \begin{eqnarray}
F^{\nu\mu}j_{s\mu}^{(2)}
&=&{-}\xi ^{\varepsilon \omega}(\omega\cdot B) \varepsilon^\nu
{+} \xi ^{\varepsilon \omega} (\varepsilon\cdot B) \omega^\nu
+\xi ^{\varepsilon\varepsilon} \varepsilon^2 E^\nu +\xi ^{\omega\omega}\omega^2 E^\nu\nonumber\\
& &  {+}\xi ^{\omega E} (E \cdot B) \omega^\nu
{-}\xi ^{\varepsilon E}  (\varepsilon\cdot E) E^\nu
+ (\xi ^{\omega B}{-} \xi ^{\omega E}) (\omega\cdot B) E^\nu \nonumber\\
& &+\xi ^{E E}  E^2 E^\nu+ (\xi ^{B B}{-}\xi ^{E B}) B^2 E^\nu  +  \xi ^{E B}( E \cdot B) B^\nu
\end{eqnarray}
Then  from the conservation law $\partial_\mu  T ^{(2)\mu\nu}=F^{\nu\mu}j_{s\mu}^{(2)}$,
we obtain the  equations that could determine or constrain these coefficients.
It is convenient to decompose these equations into three groups:  The  group I
 includes  the coefficients for the pure $\varepsilon^\mu$ and $\omega^\mu$ term in energy-momentum tensor,
\begin{eqnarray}
\mathcal{X}_{1}=0,\ \ \ \ \mathcal{X}_{2}=0,\ \ \ \ \mathcal{X}_{3}=0
\end{eqnarray}
the  group II  contains  the mixed terms between  electromagnetic field and vorticity field in energy-momentum tensor
\begin{eqnarray}
& &\mathcal{X}_{4}=-\xi^{\varepsilon\omega},\ \ \ \ \mathcal{X}_{5}=\xi^{\varepsilon\omega},\ \ \ \ \mathcal{X}_{6}=0,
,\ \ \ \ \mathcal{X}_{7}=\xi^{\varepsilon\varepsilon},\ \ \ \  \mathcal{X}_{8}=\xi^{\omega\omega}
\end{eqnarray}
and the  group III involves   the pure electromagnetic terms in energy-momentum tensor,
\begin{eqnarray}
\mathcal{X}_{9} &=& 0,\ \ \   \mathcal{X}_{10}=0,\ \ \  \mathcal{X}_{11}=\xi^{\omega E},
\ \   \ \mathcal{X}_{12}=-\xi^{\varepsilon E},\ \  \  \mathcal{X}_{13}=\xi^{\omega B}- \xi^{\omega E},\nonumber\\
 \mathcal{X}_{14}&=&\xi^{E E},
\  \ \ \mathcal{X}_{15}=\xi^{BB}-\xi^{EB},\ \ \  \mathcal{X}_{16}=\xi^{EB}
\end{eqnarray}
We note that if we know the coefficients in the energy-momentum tensor, we can directly obtain the coefficients in
the charge current from the group II or the  group III.
At second order for the chiral fermions, the energy-momentum tensor would include trace anomaly
which can lead to extra constraint identities referred as the group IV
\begin{eqnarray}
0 &=& \lambda ^{\varepsilon\varepsilon} + 3\bar \lambda ^{\varepsilon\varepsilon} + \tilde\lambda ^{\varepsilon\varepsilon},\\
0 &=& \lambda ^{\omega\omega} + 3\bar \lambda ^{\omega\omega} + \tilde\lambda ^{\omega\omega},\\
0 &=& \lambda ^{\varepsilon E} + 3\bar \lambda ^{\varepsilon E}  + 2\tilde  \lambda ^{\varepsilon E},\\
0 &=& \lambda ^{\omega B}  + 3\bar\lambda ^{\omega B} + 2\tilde\lambda ^{\omega B},\\
\tilde{C} &=& \lambda ^{E E} + 3\bar \lambda ^{E E} + \tilde\lambda ^{E E},\\
-\tilde{C} &=& \lambda ^{B B} + 3\bar \lambda ^{B B} + \tilde\lambda ^{B B}
\end{eqnarray}
From the  group I together with the first two equations in the group IV, we note that only three
coefficients are independent.  From the  naive dimension analysis, we know that  these coefficients in group I must take the form
of $T^2$.  Choosing $\lambda ^{\varepsilon\varepsilon}$, $\lambda ^{\omega\omega}$ and $\bar\lambda ^{\omega\omega}$ as independent
variables, we can obtain
\begin{eqnarray}
 \tilde\lambda_s^{\varepsilon\varepsilon}&=& 0 ,\\
 \bar\lambda_s^{\varepsilon\varepsilon}&=& -\frac{1}{3}\lambda_s^{\varepsilon\varepsilon},\\
\tilde\lambda_s^{\omega\omega}&=& -\lambda_s^{\omega\omega} -3\bar\lambda_s^{\omega\omega} ,\\
\lambda_s^{\varepsilon\omega}&=&- \frac{1}{3}\lambda_s^{\varepsilon\varepsilon} +\frac{1}{2} \lambda_s^{\omega\omega}
+ \frac{5}{2}\bar\lambda_s^{\omega\omega}
\end{eqnarray}
 Once these coefficients  have been already known, from the group II and  group IV together with the naive dimension analysis
$\lambda ^{\varepsilon E}, \bar\lambda ^{\varepsilon E}, \tilde\lambda ^{\varepsilon E}, \bar\lambda ^{\omega B},
\tilde\lambda ^{\omega B}, \lambda ^{\omega E} \propto T$, we find that $\xi^{\varepsilon\varepsilon}$,$\xi^{\omega\omega}$ and   $\xi^{\varepsilon\omega}$  in  $j^{(2)\mu}$ satisfy the following constraint
 \begin{eqnarray}
\xi_s^{\varepsilon\varepsilon} - \xi_s^{\varepsilon \omega} -\xi_s^{\omega\omega}
&=&-\frac{1}{T}\frac{\partial \bar\lambda_s^{\varepsilon\varepsilon}}{\partial \bar \mu_s}
+\frac{1}{T}\frac{\partial \bar\lambda_s^{\omega\omega}}{\partial \bar\mu_s}
+\frac{1}{T}\frac{\partial \tilde\lambda_s^{\omega\omega}}{\partial \bar\mu_s},\\
\end{eqnarray}
which indicates that only two of $\xi^{\varepsilon\varepsilon}$,$\xi^{\omega\omega}$ and
  $\xi^{\varepsilon\omega}$  are independent. Still from the group II with known  $\xi^{\varepsilon\varepsilon}$, we have
\begin{eqnarray}
\bar\lambda_s^{\varepsilon E} &=&\frac{1}{2} \left(\xi_s^{\varepsilon\varepsilon}+
\frac{1}{T}\frac{\partial \bar\lambda_s^{\varepsilon\varepsilon}}{\partial \bar \mu_s}\right),
\end{eqnarray}
which further leads to
\begin{eqnarray}
\bar \lambda_s^{\omega B} &=& \frac{1}{2}\left(\xi_s^{\varepsilon\omega } -2 \bar\lambda_s^{\varepsilon E}\right),
\end{eqnarray}
Among the other transport coefficients for  the mixed terms in energy-momentum tensor, we find only one transport coefficient is
 independent. We will choose $\lambda_s^{\omega B}$ as the independent one and  from the group II and the middle two equations
 in the trace constraint equations, we can express other coefficients as the following
\begin{eqnarray}
\tilde\lambda_s^{\omega B} &=& -\frac{1}{2}\left(\lambda_s^{\omega B} + 3 \bar\lambda_s^{\omega B}\right),\\
\tilde\lambda_s^{\varepsilon E} &=& -\frac{1}{2}\left(\lambda_s^{\omega B} +  \bar\lambda_s^{\omega B}\right),\\
\lambda_s^{\varepsilon E} &=& -\left(3\bar\lambda_s^{\varepsilon E} + 2 \tilde\lambda_s^{\varepsilon E}\right),\\
\lambda_s^{\omega E} &=& \frac{1}{2}\left(2\bar\lambda_s^{\omega B} + 4 \tilde\lambda_s^{\omega B}
-\frac{1}{T}\frac{\partial\tilde\lambda_s^{\omega\omega}}{\partial \bar\mu_s}\right)
 \end{eqnarray}
From the last equations in the  group III, it is straightforward to obtain
\begin{eqnarray}
\tilde\lambda_s^{BB} &=&-\int T \xi_s^{EB} d\bar\mu_s,\\
\bar\lambda_s^{BB} &=& - \int T \left(\xi_s^{BB}-\xi_s^{EB} \right) d\bar\mu_s,
\end{eqnarray}
where $\int T \xi ^{XX} d\bar\mu$ denotes the undetermined integral and possibly includes arbitrary functions with temperature dependence.
Then from  the  group III together with the trace anomaly in group IV, the other coefficients can be totally determined by
\begin{eqnarray}
\lambda_s^{BB} &=& -\tilde{C} - 3\bar \lambda_s^{B B} - \tilde\lambda_s^{B B},\\
\tilde\lambda_s^{EE} &=& -\frac{1}{2}
\left(\xi_s^{\omega B} - 2\tilde \lambda_s^{BB}
+ \frac{1}{T}\frac{\partial \bar\lambda_s^{\omega B}}{\partial \bar\mu_s}
+ \frac{2}{T}\frac{\partial \tilde\lambda_s^{\omega B}}{\partial \bar\mu_s}\right),\\
\bar\lambda_s^{EE} &=&- \tilde\lambda_s^{EE}-\int T \xi_s^{EE}d\bar \mu_s ,\\
\lambda_s^{EE} &=& \tilde{C}-3\bar\lambda_s^{EE}-\tilde\lambda_s^{EE},\\
\label{dTemperature-1}
\lambda_s^{EB} &=& \frac{1}{4}
\left(\xi_s^{\omega B} - 2\xi_s^{\omega E} + 4 \bar\lambda^{EE}
+ 4 \tilde \lambda^{EE} + 4 \bar\lambda_s^{BB} + 4 \tilde \lambda_s^{BB}
- 2  T\frac{\partial\tilde\lambda_s^{BB}}{\partial T}
+ \frac{1}{T}\frac{\partial \bar\lambda_s^{\omega B}}{\partial \bar\mu_s}
- \frac{2}{T}\frac{\partial \tilde\lambda_s^{\omega B}}{\partial \bar\mu_s}\right)
\end{eqnarray}
Three independent equations have not been used and remained as the constraint conditions:
\begin{eqnarray}
\label{dTemperature-2}
0&=&\bar\lambda_s^{E E} + \lambda_s^{E E}+2 \bar\lambda_s^{B B} {+ 3}\tilde\lambda_s^{B B}
- 2\lambda_s^{E B} + T \frac{\partial \bar\lambda_s^{E E}}{\partial T}
+ \frac{1}{T}\frac{\partial\tilde\lambda_s^{\varepsilon E}}{\partial \bar\mu_s}
- \frac{1}{T} \frac{\partial \tilde\lambda_s^{\omega B}}{\partial \bar\mu_s} ,\\
\label{dTemperature-3}
0&=& \bar\lambda_s^{B B} + \tilde\lambda_s^{B B} -  \lambda_s^{B B}
-T\frac{\partial \bar\lambda_s^{B B}}{\partial T}-T\frac{\partial\tilde\lambda_s^{BB}}{\partial T},\\
\label{dTemperature-4}
\xi_s^{\varepsilon E}&=& - 2\bar\lambda_s^{E E} - 2 \bar\lambda_s^{B B}
-\tilde\lambda_s^{E E} {-3}\tilde\lambda_s^{B B} + 2 \lambda_s^{E B}
+ T\frac{\partial\tilde\lambda_s^{E E}}{\partial T}
+ \frac{1}{T}\frac{\partial \bar\lambda_s^{\varepsilon E}}{\partial \bar \mu_s}
+ \frac{1}{T}\frac{\partial\tilde\lambda_s^{\varepsilon E}}{\partial \bar\mu_s}
+ \frac{1}{T} \frac{\partial \tilde\lambda_s^{\omega B}}{\partial \bar\mu_s}.
\end{eqnarray}
It should be noted that we have eliminated the partial derivative on temperature from the naive dimension analysis for the
pure $\varepsilon, \omega$ terms and mixed terms between $\varepsilon, \omega$ and $E, B$ while we kept the partial derivative
for the pure $E,B$ terms in Eqs.(\ref{dTemperature-1}-\ref{dTemperature-4}). This is because the pure  $E,B$ terms in energy-momentum
tensor could include another regularization scale due to ultraviolet divergence and  the naive dimension analysis would be broken while
there is no such complexity for the pure $\varepsilon, \omega$ terms and mixed terms. This point had been demonstrated by the direct calculation given in \cite{Yang:2020mtz}. We have checked that all these second-order results are totally consistent with the results
which had been obtained from other approaches \cite{Yang:2020mtz,Buzzegoli:2017cqy,Buzzegoli:2018wpy}.
\section{Summary}
\label{sec:summary}
When a system is in global equilibrium under electromagnetic field, only constant vorticity tensor is allowed when there is no gravity field involved. The electromagnetic and vorticity field must fulfill some constraint conditions. It turns out that these constraint conditions
can be applied to determine the non-dissipative anomalous coefficients together with the energy-momentum conservation, chiral anomaly and trace anomaly.

At zeroth order, we find that the energy-momentum conservation and charge conservation hold automatically and trace vanishing leads to the
well-known relation between the energy density and pressure. At first order, from the chiral anomaly and energy-momentum conservation, all
the coefficients can be totally determined  up to some integral constants, which is as well as what the hydrodynamic method had achieved from
the second law of thermodynamics. The trace of the energy-momentum tensor always vanishes at first order.
At second order, we find that the charge conservation
holds automatically and we cannot say anything about the transport coefficients relevant to the charge current. However we  can relate
these transport coefficients in charge current to the ones in energy-momentum tensor by using the energy-momentum conservation law and find that
once we obtain the coefficients in energy-momentum tensor, the coefficients in charge current could be derived directly.  We find that
among the coefficients  relevant to the pure vorticity tensor in energy-momentum tensor there are only three coefficients are
independent and the other four coefficients can be expressed as the linear combination of these three coefficients. We present the formulas which
express the coefficients in the mixed terms from the electromagnetic and vorticity field as the ones associated with the pure vorticity terms in energy-momentum tensor and charge current. Further we can determine the coefficients relevant to the pure electromagnetic field in the
energy-momentum tensor from the charge current associated with electromagnetic field and  the energy-momentum tensor associated with
vorticity field. All these results do not depend on any specific interactions and are very general. They are supposed to be very helpfule
 to determine the second-order anomalous transport coefficients in various chiral systems.

\begin{acknowledgments}
This work was supported in part by NSFC under Nos. 11890710, 11890713, 12175123
 and the Natural Science Foundation of Shandong Province
under No. JQ201601.
\end{acknowledgments}

\end{document}